\documentclass[11pt,a4paper]{article}
\usepackage[T1]{fontenc}
\usepackage[utf8]{inputenc}
\usepackage{authblk}

\usepackage{geometry}
\usepackage{amsfonts,amsmath,amssymb,amsthm,nicefrac}
\usepackage{ulem}
\usepackage{cancel}
\usepackage{yhmath}
\usepackage{arcs}
\usepackage{graphicx}
\usepackage{url}
\usepackage{subcaption}
\usepackage{color} 

\usepackage{url}

\title{\bf Towards parameterizing the entanglement body of a qubit pair}
\author[1,2,3]{Arsen Khvedelidze} 
\author[4]{Dimitar Mladenov}
\author[3,5]{Astghik Torosyan}

\affil[1]{A. Razmadze Mathematical Institute, Iv. Javakhishvili Tbilisi State University, Tbilisi, Georgia}
\affil[2]{Institute of Quantum Physics and Engineering Technologies, Georgian Technical University, Tbilisi, Georgia}
\affil[3]{Laboratory of Information Technologies, Joint Institute for Nuclear Research, Dubna, Russia} 
\affil[4]{Faculty of Physics, Sofia University ``St. Kliment Ohridski'', Sofia, Bulgaria}
\affil[5]{A.I. Alikhanyan National Science Laboratory (YerPhI), Yerevan, Armenia}

\date{ }

\begin{document}

\maketitle

\begin{abstract}
A method allowing to increase a computational efficiency of evaluation of non-local characteristics of a pair of qubits is described.
The method is based on the construction of coordinates on a generic section of 2-qubit's entanglement space $\mathcal{E}_{2\times2}$ represented as the direct product of an ordered 3\--dimensional simplex and the double coset 
$\mathrm{SU(2)}\times\mathrm{SU(2)}{\backslash} {\mathrm{SU(4)}}/ \mathrm{T^3}\,.$
Within this framework, the subset 
$\mathcal{SE}_{2\times2} \subset\mathcal{E}_{2\times2}$ 
corresponding to the rank\--4 separable 2-qubit states is described as a semialgebraic variety given by a system of 3rd and 4th order polynomial inequalities in eigenvalues of the density matrix, whereas  the polynomials coefficients are trigonometric functions defined over a direct product of two regular octahedra.
\end{abstract} 

\tableofcontents

\newpage

\label{sec:introduction}
\section*{Introduction}

The problem of evaluation of non-local properties of $N\--$level quantum system requires significant amount of computational resources even for a relatively small $N$. 
On the way to achieving a certain simplification there is a natural tool to use \---  the unitary symmetry 
$\mathcal{SU}(\mathcal{H}_N)$ of the underlying Hilbert space $\mathcal{H}_N$. 
Effectiveness of taking into account of the unitary symmetry can be illustrated by the following well-known example.  
Consider the adjoint action of the unitary group SU(N) on the state space 
\begin{equation}
    \mathfrak{P}_N =\{\, \varrho \in M_N(\mathbb{C}) \, | \, \varrho=\varrho^\dagger\,, \varrho \geq 0\,, \mbox{Tr}\varrho =1\,\}\,.
\end{equation} 
This action ensures the existence of a finite number of inequivalent classes and allows to evaluate the unitary invariant characteristics as functions over the corresponding  \textit{orbit space}\,\---  ${\mathfrak{P}_N}/{\mathrm{SU(N)}}\,.$ 
With a computational standpoint, the projection from state space to orbit space means an effective reduction from 
$N^2-1$ to $N-1$ independent variables.
The second example, which is more intricate and is an issue of the present report, concerns the case when an additional refine information on quantum system is available, say it is known that $N$\--dimensional system represents a union of $k$-subsystems each with $n_1, n_2, \ldots , n_k\,$ levels respectively, 
$N = \prod_{i=1}^{k}{n_i}$.  
For this system, according to the superposition principle, the Hilbert space reads 
\begin{equation}
\label{eq:TotalHS}
\mathcal{H}_{n_1\times n_2 \times \cdots \times n_k} \subseteq \mathcal{H}^{n_1}\otimes\mathcal{H}^{n_2}\otimes\cdots\otimes\mathcal{H}^{n_k}\,.
\end{equation}
As a result of the tensor product decomposition (\ref{eq:TotalHS}), the local unitary group
\begin{equation}
\label{eq:LUT} 
\mathrm{G_L = {SU(n_1)}\times{SU(n_2)}\times\cdots\times{SU(n_k)}}\subset \mathrm{SU}(N)\,\nonumber
\end{equation}
determines equivalent classes of states and the non-local characteristics of composite system are encoded in the corresponding reduced structure \cite{LindenPopescu1998}:
\begin{equation}
\label{eq:entangS}
\mathcal{E}_{\mathrm{n_1\times n_2 \times \cdots \times n_k}} = \mathfrak{P}_N/\mathrm{G_L}\,.   
\end{equation}
The quotient (\ref{eq:entangS}) accumulates complete information on the system's non-locality and deserves to be named as the \textit{``entanglement space''}.
Below we discuss the general feature of 
$\mathcal{E}_{\mathrm{n_1\times n_2 \times \cdots \times n_k}}$ 
and consider in detail a special case, the entanglement space of a pair of qubits, i.e., a binary quantum system composed from two 2-level subsystems. 
More precisely, let 
$\mathfrak{P}_4^4[\mathrm{T^3}] \subset \mathfrak{P}_4$
be a generic SU(4) stratum, consistent of the states of maximal rank, whose isotropy group is the maximal torus
$\mathrm{T^3} \subset \mathrm{SU(4)}$,  
\begin{equation}
\mathfrak{P}_4^4[\mathrm{T^3}] =\{\,
\varrho  \in 
\mathfrak{P}_4 \,
   | \, \mbox{rank}(\varrho)=4,\quad \mathrm{Iso_{{}_{SU(4)}}}[\varrho] = \mathrm{T^3}\,
    \}\,.
\end{equation}
The entanglement space $\mathcal{E}^{4}_{2\times 2}[\mathrm{T^3}]$ associated to the generic stratum $\mathfrak{P}_4^4[\mathrm{T^3}]$ is 9-dimensional and it can be locally identified with the following direct product:  
\begin{equation}
\mathcal{E}^{4}_{2\times 2} [\mathrm{T^3}] = 
\mathrm{Int}\left(\Delta_{3}\right) \times \mathbb{B}\,,
\end{equation}
where $\Delta_{3}$ is the 3-dimensional ordered simplex of eigenvalues of the density matrix 
and $\mathbb{B}$ denotes the following 6\--dimensional double coset: 
\begin{equation}
\label{eq:double}
\mathbb{B} = \mathrm{SU(2) \times SU(2) {\backslash} 
   SU(4) / {T}^3}\,.
\end{equation}
Consequently, realizing in a constructive way the projection  
$\mathfrak{P}_4  \mapsto \mathcal{E}^{4}_{2\times 2} [\mathrm{T^3}]$,  
we streamline a description of non-local quantities by reducing the number of independent variables from 15 elements of the density matrix to 9 coordinates of the entanglement space 
$\mathcal{E}^{4}_{2\times 2}[\mathrm{T^3}]$.
To realize this program, we use the parameterization of SU(4) group recently proposed in \cite{AKhMT24} and introduce a coordinate patch for almost all points of $\mathcal{E}^{4}_{2\times 2}[\mathrm{T^3}]$. 
In this picture the latter is treated locally as the direct product of 3-simplex and two copies of a regular octahedron with the edge length $2\pi\sqrt{2}.$

Apart from this, in the report the subset 
$\mathcal{SE}_{2\times2} \subset\mathcal{E}_{2\times2}$ 
corresponding to the rank\--4 \textit{separable 2-qubit states} is described as a semialgebraic variety associated to a system of 3rd and 4th order polynomial inequalities in eigenvalues of the density matrix.  
Moreover, it turns out that coefficients of the polynomials are trigonometric functions of the introduced octahedra coordinates.

\section{On the local structure of the entanglement space} 

Here we briefly formulate the generic statement concerning the representation of the entanglement space of a composite system as a stratified variety with strata given by a union of certain direct products of simplexes and double cosets.  

Let us consider stratification of the state space $\mathfrak{P}_N$ owing to the unitary symmetry.
Action of the group G = SU(N) on the state space $\mathfrak{P}_N$ ensures the equivalence relation 
$\varrho_1 \stackrel{\mathrm{SU(N)}}{\sim} \varrho_2$ 
providing its stratification with the strata enumerated by the state's isotropy groups 
$[\mathrm{H_\alpha}] \subset \mathrm{G}$\,: 
\begin{equation}
\label{eq:OrbitStrat}
\mathfrak{P}_N = \bigcup_{
\mathrm{G\--orbit~type}}\, \mathfrak{P}_{[\mathrm{H_\alpha}]}\,.
\end{equation} 

\textbf{Proposition I.}
\textit{
In the vicinity of state with the isotropy group $\mathrm{H_\alpha}$\,,
the components $\mathfrak{P}_{[\mathrm{H_\alpha}]}$ can be represented as the following direct product of two factors: 
\begin{equation}
\label{eq:Pstrata}
\mathfrak{P}_{[\mathrm{H_\alpha}] }  = \Delta_{N}^{(\alpha)}\times G/\mathrm{H_\alpha}\,,
\end{equation}
where $\Delta_N^{(\alpha)}$ is a subset of the ordered simplex $\Delta_{N-1}$ of eigenvalues of the states with a given algebraic degeneracy corresponding to the group $\mathrm{H_\alpha}$. 
}

From this proposition it follows, that for the states with a non-degenerate spectrum, the isotropy group is the maximal torus, $\mathrm{H_1=T^{N-1}}$, and $\Delta_N^{(1)}$ is the interior of the simplex $\Delta_{N-1}$. 
Similarly, for all isotropy subgroups forming the ordered hierarchy, there will be corresponding subsets of these simplexes \--- the certain facets of the large simplex, see e.g. details in  \cite{Bengtsson_Zyczkowski_2006}.

Using (\ref{eq:Pstrata}) and taking into account a double coset decomposition of SU(N), we arrive at the entanglement space decomposition.
\footnote{It is worth to emphasize that the resulting decomposition is valid only locally.
Analysis of the bundle structure of the entanglement space goes far beyond of the present report.}
To avoid combinatorial details, we write down only decomposition of a bipartite $(N=N_AN_B)$ entanglement space:  
\begin{equation}
\label{eq:entbip}
\mathcal{E}_{{}_{N_A\times N_B}} 
= \bigcup_{\mathrm{orbit~ type\, [\mathrm{H_\alpha}]}} \, \mathcal{E}_{{}_{N_A\times N_B}}[\mathrm{H_\alpha}]\quad  = \bigcup_{\mathrm{orbit~ type\, [H_\alpha]}} \,\Delta_{N}^{(\alpha)}\times \mathrm{G_L\backslash{G}}/\mathrm{H_\alpha}\,.
\end{equation}
Bearing in mind (\ref{eq:entbip}), we now  consider in detail a 2-qubits case.

\section{Example: 2-qubit entanglement space} 

In this section we exemplify the local structure of the entanglement space considering a pair of qubits. 

\subsection{\sf Density matrix parameterization. \quad}
There are different ways to para\-me\-te\-rize a density matrix of a pair of qubits. 
Mostly used form of the density operator is characterized by two Bloch 3-vectors $a,b$ of partially reduced density matrices of qubits ``A''  and ``B'' respectively, and 9 real coefficients $c_{ij}$ being elements of $3\times 3$ correlation matrix: 
\begin{equation}
\label{eq:matrf}
\varrho = \frac{1}{4}\,\mathbb{I}_4+ \frac{\imath}{2}\,
\sum_{i=1}^3a_i \sigma_{i0}+
\frac{\imath}{2}\,
\sum_{i=1}^3\,b_i \sigma_{0i} +
\frac{\imath}{2}\,
\sum_{i,j=1}^3\,c_{ij}\sigma_{ij}\,. 
\end{equation}
In (\ref{eq:matrf}) we use the Fano basis of the Lie algebra $\mathfrak{su}(4)$
\begin{equation}
\sigma_{i0}= \frac{1}{2\imath}\,
\sigma_i\otimes \mathbb{I}_2\,,\quad
\sigma_{0i}=\frac{1}{2\imath}\,
\mathbb{I}_2\otimes\sigma_i\,,
\quad 
\sigma_{ij}= \frac{1}{2\imath}\,
\sigma_i\otimes \sigma_j\,, \quad 
i,j =1,2,3\,. 
\end{equation} 
Another method to represent a 2-qubit  mixed state is to use the Singular Value Decomposition (SVD) of the density matrix $\varrho$.
Introducing one index notations for the basis elements: 
\begin{equation} 
\boldsymbol{\lambda}=\{ \sigma_{10}, \sigma_{20}, \sigma_{30}, \sigma_{01}, \sigma_{02}, \sigma_{03}, \sigma_{11}, \sigma_{12}, \sigma_{13}, \sigma_{21}, \sigma_{22}, \sigma_{23}, \sigma_{31}, \sigma_{32}, \sigma_{33} \}\,,
\end{equation}
we can write down SVD of the density matrix as 
\begin{equation}
\label{eq:SVDdens}
\varrho = U\mathrm{diag}(r_1, r_2, r_3, r_4)U^{\dagger} = \frac{1}{4}\, \mathbb{I}_4 + \frac{\imath}{2}\, U\bigg(x\,\lambda_3 + y\,\lambda_6 + z\,\lambda_{15} \bigg) U^\dagger\,,
\end{equation}
where the diagonal factor of SVD is expanded over the basis elements of the Cartan subalgebra of $\mathfrak{su}(4)$ algebra.
The expansion coefficients $x, y, z$ are real numbers parameterizing the density matrix eigenvalues $(r_1, r_2, r_3, r_4)$: 
\begin{eqnarray}
&& r_1 = \frac{1}{4}(1 + x + y + z)\,, \quad 
r_2 = \frac{1}{4}(1 + x - y - z)\,,\\ 
&& r_3 = \frac{1}{4}(1 - x + y - z)\,, \quad 
r_4 = \frac{1}{4}(1 - x - y + z)\,.
\end{eqnarray} 
Considering a maximal rank non-degenerate eigenvalues 
chosen in decreasing order, $1> r_1> r_2> r_3> r_4>0$\,, we identify  the unitary matrix $U$ as  an element of the coset $U \in \mathrm{SU(4)/T^3}\,.$ 

\bigskip 
\subsection{\sf Generic  $\mathrm{SU(2)\times SU(2)}$\--orbit  representative. \quad} 
The local unitary group $\mathrm{G_L}$ for two qubits is the subgroup 
$\mathrm{SU(2)\times SU(2) \subset SU(4)}$.
Its generic orbit $\mathcal{O}_{d=6}(\varrho)$ of a state 
$\varrho \in \mathfrak{P}_4$
with a non-degenerate spectrum is 6-dimensional.  
In order to parameterize this orbit, we use in (\ref{eq:SVDdens}) the following factorization of $\mathrm{SU(4)}$ \cite{AKhMT24}: 
\begin{equation}
\label{eq:su4}
\mathrm{SU(4)= K \mathcal{A} T^3}\,, 
\end{equation}
where $\mathrm{K = SU(2)\times SU(2)}$,
$\mathrm{T^3}$ is the maximal 3\--torus subgroup, 
and the $\mathcal{A}$-factor represents the product of two exponents, 
$\mathcal{A} = \exp{\mathfrak{a}} \exp{\mathfrak{a}^\prime}$, where 
\begin{equation}
\mathfrak{a} = \alpha_1\lambda_1+ \alpha_2\lambda_4+ \alpha_3\lambda_7\,, \quad
\mathfrak{a}^\prime = \beta_1\lambda_9+ \beta_2\lambda_{11}+ \beta_3\lambda_{13}\,.
\end{equation}
The real triplets $\boldsymbol{\alpha}=(\alpha_1, \alpha_2, \alpha_3)$ and 
$\boldsymbol{\beta} = (\beta_1, \beta_2, \beta_3)$ are coordinates of two copies of an octahedron with the edge length $2\pi \sqrt{2}$. 
As a result, we got convinced that any non-degenerate rank-4 state is $\mathrm{G_L}$\--equivalent to the following 9-parameter representative matrix $\varrho_{d=6}$:  
\begin{equation}
\label{eq:6repres}
\varrho_{d=6}=\mathcal{A}
\begin{pmatrix}
 r_1 & 0 & 0& 0\\
 0 & r_2 & 0 &0\\
 0  & 0 & r_3& 0\\
 0  & 0 & 0& r_4
\end{pmatrix}
\mathcal{A}^\dagger\,.
\end{equation} 
Within this representative of the generic $\mathrm{G_L}$\--orbits in $\mathfrak{P}_4$, we are able to make certain conclusions on the local structure of the entanglement space $\mathcal{E}_{2\times 2}$.

\textbf{Proposition II.}
\textit{
In the vicinity of a state $\varrho$ consisting of maximal rank states with a non-degenerate spectrum, 
the entanglement space admits the representation:
\begin{equation}
\label{eq:Ebody}
\mathcal{E}^4_{{}_{2\times2}} = \mathrm{Int}\left(
\Delta_{3}\right) \times \mathrm{O}_h\times\mathrm{O}_h \,,
\end{equation}
where the factor 
$\mathrm{Int} \left(\Delta_{3}\right)$ denotes the interior of the ordered 3-simplex, 
and $\mathrm{O}_h$ is regular octahedron with edge $2\pi\sqrt{2}\,.$ 
}
\bigskip 

\section{Algebraic form of a pair of qubits's mixed states separability}

In what follows, we analyze  the separability problem of mixed states of a pair of qubits.
Consider the partial transforms of a density matrix $\varrho $ of 2-qubits defined as: 
\begin{eqnarray}
  \varrho \to \varrho^{T_B}
    =(\mathbb{I}_{2}\otimes\mathrm{T}) \varrho \,,\qquad 
    \varrho \to \varrho^{T_A}
    =(\mathrm{T}\otimes\mathbb{I}_{2}) \varrho\,,
\end{eqnarray}
where $\mathrm{T}$ is transposition  in a subsystem.
The state space $\mathfrak{P}_{4}$ is not invariant under $\mathrm{T}^{A}(\mathrm{T}^{B})$ 
mappings, but there is a subset 
$\mathfrak{S}_4 \subset \mathfrak{P}_4$, named as the space of 
Positive under Partial Transposition (PPT) states, which is an  invariant subspace of $\mathfrak{P}_{4}$. 
The subset $\mathfrak{S}_4$
is of a special interest due to the 
Peres-Horodecki observation 
\cite{Peres1996,HorodeckiHorodecki99} 
that 2-qubit mixed  PPT states 
are \textit{separable} as well, i.e., they admit 
the convex decomposition:
 \begin{equation}
\varrho =\sum_{k} \omega_k \varrho_1^k\otimes\varrho_2^k\,, \qquad \sum_{k}\omega_k =1,\quad \omega_k >0\,,
\end{equation}
where $\varrho^k_1 $ and $\varrho^k_2$ are  the density matrices  of individual qubits. 
Otherwise the states are \textit{entangled} \cite{Werner}.

\subsection{\sf Separability condition in state space coordinates. \quad}
Since the semi-positivity of arbitrary density matrices 
is equivalent to the semi-positivity of the coefficients of the characteristic polynomial
\begin{equation}
\det(x -\varrho) =x^4 - x^3 +S_2(\varrho)x^2 -
S_3(\varrho)x + S_4(\varrho)\,,
\end{equation}
the subset  $\mathfrak{S}_4$ is determined by the  non-negativity of 3rd and 4th order coefficients
\footnote{Note that $S_2(\varrho)= S_2(\varrho^{T_B})$.}
\begin{equation}
\label{eq:Stb}
\mathfrak{S}_4: \ \{\,
S_3(\varrho^{T_B}) \geq 0, \, 
    \quad
    S_4(\varrho^{T_B}) \geq 0\, \}\,.
\end{equation}
Explicitly inequalities (\ref{eq:Stb}) have been represented in terms of 15 elements of the density matrix 
$\varrho$, the correlation matrix $C$ and the Bloch vectors $\boldsymbol{a}, \boldsymbol{b}$ of individual qubits\cite{GKP2011}: 
\begin{eqnarray}
\label{eq:Sep} 
&&0 \leq S_3+\frac{1}{4}\, \det||C|| \leq  \frac{1}{16}\,,\\
&& 0\leq S_4 + \frac{1}{16}\, \det||M|| \leq  
\frac{1}{256}\,,  
\end{eqnarray}
where $M_{ij} = C_{ij}-a_ib_j$ denotes the Schlienz-Mahler $3\times 3$ matrix.

\subsection{\sf Separability condition in the entanglement space coordinates. \quad} Now noting that the separability conditions are $\mathrm{SU(2)}\times \mathrm{SU}(2)$\--invariant, we are able to rewrite  separability conditions (\ref{eq:Sep}) in terms of 
the representative state $\varrho_{d=6}\,.$ 
Our calculations of  the correlation matrix determinant give: 
\begin{equation}
\det||C||= z\left(p_{201}\left(x^2+y^2\right) + p_{111}\,x y\right)\,, 
\end{equation}
where 
\begin{flalign}
p_{201} &= \frac{1}{4}\sin \left(2 \alpha _3\right) \sin \left(2 \beta _2\right) \cos \left(\beta _1\right) \cos \left(\beta _3\right)\,, \\ 
p_{111} &= -\Big(\sin ^2\left(\alpha _3\right) \cos ^2\left(\beta _2\right) \left(\cos ^2\left(\beta _1\right)+\sin ^2\left(\beta _1\right) \cos ^2\left(\beta _3\right)\right) +&&\\ 
\nonumber
&\cos ^2\left(\alpha _3\right) \sin ^2\left(\beta _2\right) \cos ^2\left(\beta _1\right) \cos ^2\left(\beta _3\right) + \sin ^2\left(\beta_1\right)
\sin^2\left(\beta_3\right) \cos^2\left(\beta_2\right)\Big). &&
\end{flalign}
It is worth noting that the coefficients $p_{ijk}$ demonstrate certain symmetry since they depend only on the 4 coordinates of the octahedra. 
In order to find the expression for 
Schlienz-Mahler matrix determinant, we use the following identity: 
\begin{equation}
\det||M||= \det||C||- \frac{1}{2}\, C^{(112)}\,,
\end{equation} 
relating the determinats of $M$ and
$C$ matrices via the 4th order Quesne homogeneous  $\mathrm{SU(2)\times SU(2)}$\--invariant  polynomial \cite{Quesne1976}: 
\begin{equation}
C^{(112)} =
\epsilon_{ijk}\epsilon_{\alpha\beta\gamma}a_i b_\alpha
    c_{j\beta}c_{k\gamma}\,.
\end{equation}
The calculation of Quesne polynomials is cumbersome, but their structure in SVD coordinates can be easily found out, particularly, 
\begin{equation}
\label{eq:112}
 C^{(112)} =\sum_{i_1+i_2+i_3=4 }\, 
    p_{i_1 i_2 i_3}(\boldsymbol{\alpha,\boldsymbol{\beta}})\,x^{i_1}y^{i_2}z^{i_3}\,.
\end{equation}
In general, a homogeneous fourth-degree polynomial in three variables contains 15 coefficients. 
But our calculations show that among the 15 coefficients $p_{ijk}$ in (\ref{eq:112}) only 9 are non-vanishing. Moreover, again it turns that all of them are functions of just 4 variables, one coordinate $\alpha_3$ of the first octahedron and all three coordinates $\boldsymbol{\beta}$ of the second octahedron. We omit 
bulky formulae  for all coefficients and give here as an example only one typical expression: 
\begin{eqnarray}
p_{022}&=&
\frac{1}{8} 
\cos^2\left(\alpha _3\right) \cos^2\left(\beta_1\right) \left(\cos \left(2 \left(\alpha_3-
\beta_1\right)\right)+\cos \left(2 \left(\alpha _3+\beta _1\right)\right)\right.\nonumber \\
&+&
4 \cos \left(2 \beta _3\right) \left(\cos \left(2 \beta _2\right) \left(1-\cos ^2\left(\alpha _3\right) \cos \left(2 \beta _1\right)\right)+\sin ^2\left(\alpha _3\right)\right)\nonumber \\
&-&
\left.4 \sin ^2\left(\alpha _3\right) \cos \left(2 \beta _2\right)+2 \cos \left(2 \beta _1\right)-4\right)\,.
\end{eqnarray}

\label{sec:summary}
\section{Summary}

In the present note we discussed the stratified structure of the entanglement space of two qubits $\mathcal{E}_{2\times 2}$ and described its subset $\mathcal{SE}_{2\times 2}$ corresponding to the separable states as 7-dimensional semialgebraic variety. 
In the forthcoming  publication we will extend this formulation for the complementary part of the entanglement space $\mathcal{E}_{2\times 2}$  using the polynomial basis for $\mathrm{SU(2)\times SU(2)}$ invariants (see e.g. \cite{GKP2009} and references therein) rewritten analogously in terms of eigenvalues of 2-qubit density matrix and coordinates of two octahedra.


\end{document}